\documentclass[twocolumn,showpacs,amsmath,amssymb]{revtex4}

\usepackage{graphicx}
\usepackage{graphics}
\usepackage{dcolumn}
\usepackage{bm}
\def\rr{{{\bf r}}}
\def\rrp{{{\bf r}^\prime}}

\begin{document}

\title{Exact exchange optimized effective potential and self-compression of\\ stabilized jellium clusters}

\author{M. Payami}
\affiliation{Center for Theoretical Physics and Mathematics,
Atomic Energy Organization of Iran, P. O. Box 11365-8486,
Tehran-Iran }
\date{\today}
\begin{abstract}
In this work, we have used the exchange-only optimized effective
potential in the self-consistent calculations of the density
functional Kohn-Sham equations for simple metal clusters in
stabilized jellium model with self-compression. The results for
the closed-shell clusters of Al, Li, Na, K, and Cs with $N=$2, 8,
18, 20, 34, and 40 show that the clusters are $3\%$ more
compressed here than in the local spin density approximation. On
the other hand, in the LSDA, neglecting the correlation results in
a contraction by $1.4\%$.
\end{abstract}

\maketitle

\section{Introduction}\label{sec1}
The Kohn-Sham (KS)\cite{KS65} density functional theory
(DFT)\cite{HK64} is one of the most powerful techniques in
electronic structure calculations. However, the exact form of the
exchange-correlation functional is still unknown and, in practice,
one must use approximations. The accuracy of the predictions of
the properties depends on how one approximates this functional.
The simplest one is the local spin density approximation (LSDA) in
which one uses the properties of the uniform electron gas locally
\cite{KS65}. This approximation is in principle appropriate for
systems in which the variations of the spin densities $n_\sigma$
are sufficiently slow. For finite systems and surfaces which are
highly inhomogeneous, the generalized gradient approximation
(GGA)\cite{Perdew96} is more appropriate. In spite of the success
of the LSDA and GGA, it is observed that in some cases these
approximations fail to predict even qualitatively correct
behaviors\cite{Schmid,Varga,Leung,Dufek}. On the other hand,
appropriate self-interaction corrected versions of these
approximations are observed to lead to correct
behaviors\cite{Dufek,Engel93}. These observations motivates one to
use functionals in which the self-interaction contribution is
removed exactly. One of the functionals, which satisfies this
constraint, is the exact exchange (EEX) orbital dependent
functional. Using the EEX functional leads to the correct
asymptotic behavior of the KS potential as well as to correct
results for the high density limit in which the exchange energy is
dominated \cite{Perdew01}. Although neglecting the correlation
effects in orbital dependent functionals fails to reproduce the
dispersion forces such as the van der Waals
forces\cite{Engel99,Magyar04}, the EEX in some respects is
advantageous over the local and semi-local
approximations\cite{Magyar04,Kummel04}. To obtain the local
exchange potential from the orbital dependent functional, one
should solve the optimized effective potential (OEP) integral
equation. Recently, K\"ummel and Perdew
\cite{KummelPRL03,KummelPRB03} have invented an iterative method
which allows one to solve the OEP integral equation accurately and
efficiently even for three dimensional systems. This method is
used in this work.

To simplify the cluster problem, one notes that the properties of
alkali metals are dominantly determined by the delocalized valence
electrons. In these metals, the Fermi wavelengths of the valence
electrons are much larger than the metal lattice constants and the
pseudo-potentials of the ions do not significantly affect the
electronic structure. This fact allows one to replace the discrete
ionic structure by a homogeneous positive charge background which
is called jellium model (JM). In its simplest form, one applies
the JM to metal clusters by replacing the ions of an $N$-atom
cluster with a sphere of uniform positive charge density and
radius $R=(zN)^{1/3}r_s$, where $z$ is the valence of the atom and
$r_s$ is the bulk value of the Wigner-Seitz (WS) radius for
valence electrons\cite{Payami99,Payami01,Payami-PSS04}. Assuming
the spherical geometry is justified only for closed-shell clusters
which is the subject in this work. However, it is a known fact
that the JM has some drawbacks\cite{Lang70,Ashcroft67}. The
stabilized jellium model (SJM) in its original form\cite{Perdew90}
was the first attempt to overcome the deficiencies of the JM and
still keeping the simplicity of the JM. Application of the SJM to
simple metals and metal clusters has shown significant
improvements over the JM results\cite{Perdew90}. However, for
small metal clusters the surface effects are important and the
cluster is self-compressed due to its surface tension. This effect
has been successfully taken into account by the SJM which is
called SJM with self-compression
(SJM-SC)\cite{Perdew93,Payami-CJ04}. Application of the
LSDA-SJM-SC to neutral metal clusters has shown that the
equilibrium $r_s$ values of small clusters are smaller than their
bulk counterparts and approaches to it for very large clusters.
This trend is consistent with the results of {\it ab. initio.}
calculations\cite{Rothlisberger,Payami-PSS01}.

In this work we have used the EEX-SJM-SC to obtain the equilibrium
sizes and energies of closed-shell neutral $N$-electron clusters
of Al, Li, Na, K, and Cs for $N$=2, 8, 18, 20, 34, and 40 (for Al,
$N$ = 18 corresponds to Al$_6$ cluster and other values do not
correspond to a real Al$_n$). In order to have an estimate for the
self-interaction effects, we have repeated the calculations for
exchange-only local spin density approximation (x-LSDA) in which
the spin-polarized version of the Dirac form, $E_x=c_x\int d\rr
\;n^{4/3}$, is used. Comparison of the results shows that (except
for $N=40$ in Al case) the relation $\bar r_{EEX}<\bar
r_{x-LSDA}<\bar r_{LSDA}$. The organization of this paper is as
follows. In section \ref{sec2} we explain the calculational
schemes. Section \ref{sec3} is devoted to the results of our
calculations and finally, we conclude this work in section
\ref{sec4}.
\section{Calculational schemes}\label{sec2}
In this section we first explain how to implement the exact
exchange in the SJM, and then will explain the procedure for the
OEP calculations.

\subsection{Exact exchange stabilized jellium model}
 As in the original SJM\cite{Perdew90}, here the Ashcroft empty core
pseudo-potential\cite{Ashcroft66} is used for the interaction of
an ion of charge $z$ with an electron at a relative distance $r$:
\begin{equation}\label{eq1}
  w(r)=\left\{\begin{array}{ccc}
    -2z/r &,& (r>r_c) \\
      0   &,& (r<r_c) \
  \end{array}\right.
\end{equation}
The core radius, $r_c$, will be fixed by setting the pressure of
the bulk system equal to zero. In the EEX-SJM, the average energy
per valence electron in the bulk with density $n$ is given by
\begin{equation}\label{eq2}
  \varepsilon(n)=t_s(n)+\varepsilon_x(n)+\bar{w}_R(n,r_c)+\varepsilon_M(n),
\end{equation}
with
\begin{equation}\label{eq3}
  t_s(n)=c_kn^{2/3},
\end{equation}
\begin{equation}\label{eq4}
  \varepsilon_x(n)=c_xn^{1/3},
\end{equation}
\begin{equation}\label{eq5}
  c_k=\frac{3}{5}(3\pi^2)^{2/3},\;\;\;\;\;\;
  c_x=\frac{3}{2}(3/\pi)^{1/3}.
\end{equation}

 All equations throughout this paper are expressed in Rydberg
atomic units. Here $t_s$ and $\varepsilon_x$ are the kinetic and
exchange energy per particle, respectively. $\bar{w}_R$ is the
average value of the repulsive part of the pseudo-potential
($\bar{w}_R=4\pi nr_c^2$), and $\varepsilon_M$ is the average
Madelung energy. Demanding zero pressure for the bulk system at
equilibrium yields:
\begin{equation}\label{eq6}
 \left\{2t_s(n)+\varepsilon_x(n)+12\pi nr_c^2+
 \varepsilon_M(n)\right\}_{n=n^B}=0.
\end{equation}
Solution of this equation for $r_c$ gives
\begin{equation}\label{eq7}
  r_c(r_s^B)=\frac{(r_s^B)^{3/2}}{3}\left\{-2t_s(r_s)-
  \varepsilon_x(r_s)-\varepsilon_M(r_s)\right\}_{r_s=r_s^B}^{1/2}.
\end{equation}

\begin{figure}[htb]
\includegraphics[width=9cm]{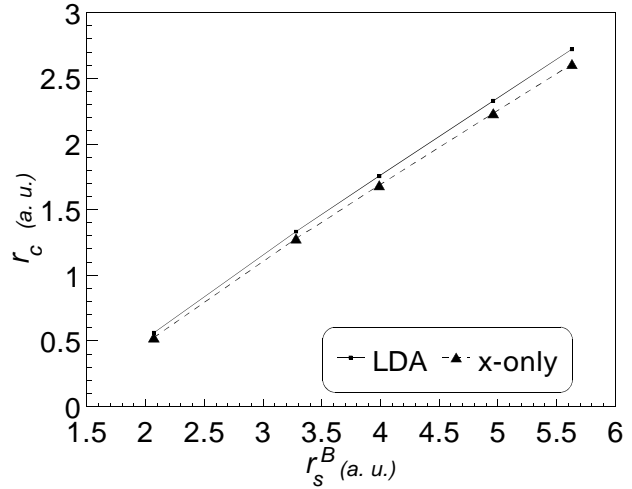}
\caption{\label{fig1} Pseudo-potential core radii in atomic units
for different $r_s^B$ values.}
\end{figure}

In Fig. \ref{fig1} we have plotted the core radii for different
values of $r_s^B$ which assume 2.07, 3.28, 3.99, 4.96, and 5.63
for Al, Li, Na, K, and Cs, respectively. The result is compared
with the case in which the correlation energy is also incorporated
(see Eq.(26) of Ref.\cite{Perdew90}). As is seen, to stabilize the
bulk system in the EEX case, the core radii assume smaller values.

 As in the original SJM\cite{Perdew90} (but in the absence of the correlation
energy component), at equilibrium density we have
\begin{equation}\label{eq8}
  \langle\delta
  v\rangle_{WS}=-\frac{1}{3}[2t_s(n^B)+\varepsilon_x(n^B)].
\end{equation}
Here, $\langle\delta  v\rangle_{WS}$ is the average of the
difference potential over the WS cell and the difference
potential, $\delta v$, is defined as the difference between the
pseudo-potential of a lattice of ions and the electrostatic
potential of the jellium positive background. Once the values of
$\langle\delta v\rangle_{WS}$ and $r_c$ as functions of $r_s^B$
are found, the EEX-SJM total energy of a cluster becomes
\begin{eqnarray}\label{eq9}
  \nonumber E_{EEX-SJM}[n_\uparrow,n_\downarrow,r_s,r_s^B]=
  E_{EEX-JM}[n_\uparrow,n_\downarrow,r_s] \\
  \nonumber +(\varepsilon_M+\bar{w}_R)\int d\rr\; n_+(\rr)\;\;\;\;\;
  \\
  +\langle\delta v\rangle_{WS}\int d\rr\;\Theta(\rr)[n(\rr)-n_+(\rr)].
\end{eqnarray}
Here,
\begin{eqnarray}\label{eq10}
\nonumber
E_{EEX-JM}[n_\uparrow,n_\downarrow,r_s]=T_s[n_\uparrow,n_\downarrow]+
   E_x[n_\uparrow,n_\downarrow]\\
   +\frac{1}{2}\int d\rr\;\phi([n,n_+];\rr)\,[n(\rr)-n_+(\rr)],
\end{eqnarray}
\begin{equation}\label{eq11}
  E_x=\sum_{\sigma=\uparrow,\downarrow}\sum_{i,j=1}^{N_\sigma}
  \int d\rr \, d\rrp
  \frac{\phi_{i\sigma}^*(\rr)\phi_{j\sigma}^*(\rrp)
  \phi_{j\sigma}(\rr)\phi_{i\sigma}(\rrp)}{\mid\rr-\rrp\mid},
\end{equation}
\begin{equation}\label{eq12}
  \phi([n,n_+];\rr)=2\int
  d\rrp\frac{[n(\rrp)-n_+(\rrp)]}{\mid\rr-\rrp\mid},
\end{equation}
\begin{equation}\label{eq13}
  n(\rr)=\sum_{\sigma=\uparrow,\downarrow}\sum_{i=1}^{N_\sigma}
  \mid\phi_{i\sigma}(\rr)\mid^2,
\end{equation}
\begin{equation}\label{eq14}
  n_+(\rr)=n\theta(R-r);\;\;\;\;\;\;n=\frac{3}{4\pi r_s^3}.
\end{equation}

 To obtain the equilibrium size and energy of an $N$-atom
cluster in EEX-SJM-SC, we solve the equation
\begin{equation}\label{eq15}
  \left.\frac{\partial}{\partial
  r_s}E(N,r_s,r_c)\right|_{r_s=\bar{r}_s(N)}=0,
\end{equation}
where $N$ and $r_c$ are kept constant and $E$ is given by Eq.
(\ref{eq9}). The preocedure for the x-LSDA is the same as above
except for that the Dirac exchange energy must be used.
\subsection{The OEP equations}
 K\"ummel and Perdew\cite{KummelPRB03} have proved, in a simple way, that the OEP
integral equation is equivalent to
\begin{equation}\label{eq16}
  \sum_{i=1}^{N_\sigma}\psi_{i\sigma}^*(\rr)\phi_{i\sigma}(\rr)+c.c.=0.
\end{equation}
$\phi_{i\sigma}$ are the self-consistent KS orbitals and
$\psi_{i\sigma}$ are orbital shifts. The self-consistent orbital
shifts and the local exchange potentials are obtained from the
iterative solutions of inhomogeneous KS equations. Taking
spherical geometry for the jellium background and inserting
\begin{equation}\label{eq27}
  \phi_{i\sigma}(\rr)=\frac{\chi_{i\sigma}(r)}{r}Y_{l_i,m_i}(\Omega),
\end{equation}
and
\begin{equation}\label{eq28}
  \psi_{i\sigma}(\rr)=\frac{\xi_{i\sigma}(r)}{r}Y_{l_i,m_i}(\Omega),
\end{equation}
in to the inhomogeneous KS equation (Eq.(21) of
Ref.\cite{KummelPRB03}) one obtains\cite{Payami05}

\begin{equation}\label{eq29}
  \left[\frac{d^2}{dr^2}+\varepsilon_{i\sigma}-v_{eff\sigma}(r)-\frac{l_i(l_i+1)}{r^2}\right]\xi_{i\sigma}(r)
  =q_{i\sigma}(r).
\end{equation}

Here, $\varepsilon_{i\sigma}$ are the KS eigenvalues and

\begin{equation}\label{eq19}
 v_{eff\sigma}(\rr)=v(\rr)+v_H(\rr)+v_{x\sigma}(\rr),
\end{equation}
\begin{equation}\label{eq20}
  v_H(\rr)=2\int d\rr\,\frac{n(\rrp)}{\mid\rr-\rrp\mid}.
\end{equation}

The right hand side of Eq. (\ref{eq29}) can be written as

\begin{equation}\label{eq30}
  q_{i\sigma}(r)=q_{i\sigma}^{(1)}(r)+q_{i\sigma}^{(2)}(r),
\end{equation}
with
\begin{equation}\label{eq31}
  q_{i\sigma}^{(1)}(r)=\left[v_{xc\sigma}(r)-\bar
  v_{xci\sigma}+\bar u_{xci\sigma}\right]\chi_{i\sigma}(r),
\end{equation}
and
\begin{eqnarray}\label{eq21}
q_{i\sigma}^{(2)}(r)=2\sum_{j=1}^{N_\sigma}\sum_{l=|l_i-l_j|}^{l_i+l_j}\frac{4\pi}{2l+1}\chi_{j\sigma}(r)B_\sigma(i,j,l;r)
\nonumber\\
\times\overline{\left[I(l_jm_j,l_im_i,lm_j-m_i)\right]^2}.
\end{eqnarray}

The quantities $B$ and $I$ in Eq. (\ref{eq21}) are defined as
\begin{eqnarray}\label{eq22}
  B_\sigma(i,j,l;r)=\int_{r^\prime=0}^rdr^\prime\chi_{i\sigma}(r^\prime)\chi_{j\sigma}(r^\prime)\frac{{r^\prime}^l}{r^{l+1}}
  \nonumber\\
  +\int_{r^\prime=r}^\infty dr^\prime\chi_{i\sigma}(r^\prime)\chi_{j\sigma}(r^\prime)\frac{r^l}{{r^\prime}^{l+1}}
\end{eqnarray}
\begin{equation}\label{eq23}
  I(l_jm_j,l_im_i,lm)=\int d\Omega\;
  Y_{l_jm_j}^*(\Omega)Y_{l_im_i}(\Omega)Y_{lm}(\Omega),
\end{equation}
and the bar over $I^2$ implies average over $m_i$ and $m_j$. Also,
the expression for $\bar u_{xi\sigma}$ reduces to
\begin{eqnarray}\label{eq21-1}
\bar
u_{xi\sigma}=-2\sum_{j=1}^{N_\sigma}\sum_{l=|l_i-l_j|}^{l_i+l_j}\frac{4\pi}{2l+1}
\overline{\left[I(l_jm_j,l_im_i,lm_j-m_i)\right]^2}\nonumber \\
\times\int_0^\infty dr\;\chi_{i\sigma}(r)\chi_{j\sigma}(r)
B_\sigma(i,j,l;r).\;\;\;
\end{eqnarray}

The procedure for the self-consistent iterative solutions of the
OEP equations is explained in Refs.\cite{KummelPRB03,Payami05}.

\begin{figure}[htb]
\includegraphics[width=9cm]{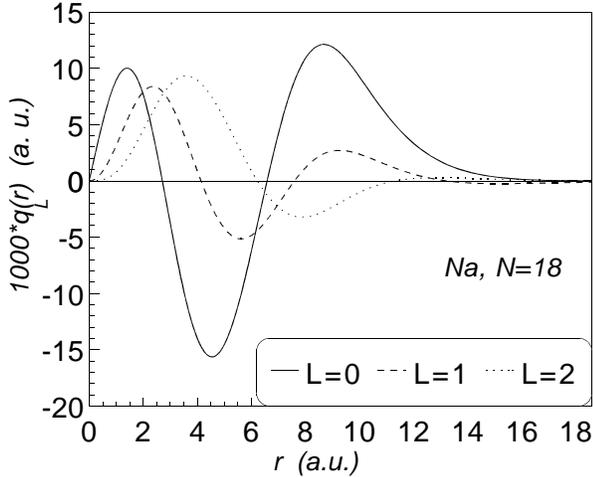}
\caption{\label{fig2} Right hand side of Eq. (\ref{eq29}) for the
self-consistent equilibrium size of Na$_{18}$.}
\end{figure}

\begin{figure}[htb]
\includegraphics[width=9cm]{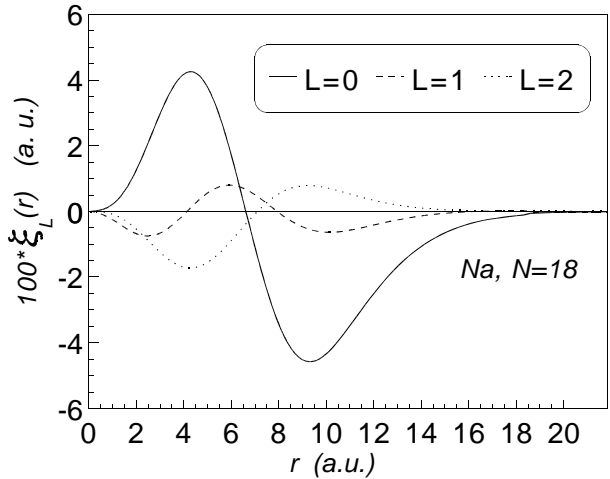}
\caption{\label{fig3} Orbital shifts in atomic units for the
self-consistent equilibrium size of Na$_{18}$.}
\end{figure}

In Fig. \ref{fig2}, the self-consistent source terms
$q_{i\sigma}(r)$ of Eq. (\ref{eq29}) are plotted for the
equilibrium size of Na$_{18}$ cluster. The corresponding orbital
shifts $\xi_{i\sigma}(r)$ are shown in Fig.\ref{fig3}.

\section{Results and discussion}\label{sec3}
We have used the EEX-SJM-SC to obtain the equilibrium sizes and
energies of closed-shell 2, 8, 18, 20, 34, and 40-electron neutral
clusters of Al, Li, Na, K, and Cs.

\begin{figure}[htb]
\includegraphics[width=9cm]{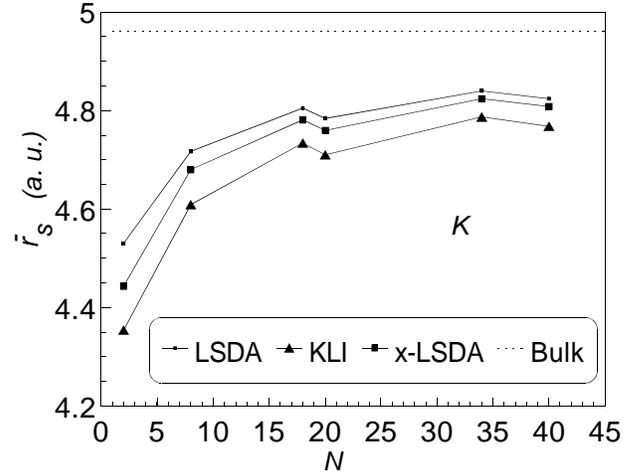}
\caption{\label{fig4} Equilibrium $r_s$ values of K$_N$ clusters
for different sizes. The dotted line is $r_s^B$. KLI and OEP
predict smaller sizes for the clusters than the x-LSDA and LSDA.}
\end{figure}

In Table \ref{table1} we have listed the equilibrium $r_s$ values,
total energies and exchange energies. As is seen, the equilibrium
$r_s$ values of the clusters are almost the same up to 3 decimals
for the KLI and OEP schemes whereas, there are significant
differences between the OEP, x-LSDA, and LSDA values. As an
example, we have plotted the equilibrium $r_s$ values of the
closed-shell K$_N$ clusters in Fig. \ref{fig4}. It shows that the
LSDA predicts larger cluster sizes than the x-LSDA and OEP.

\begin{figure}[htb]
\includegraphics[width=9cm]{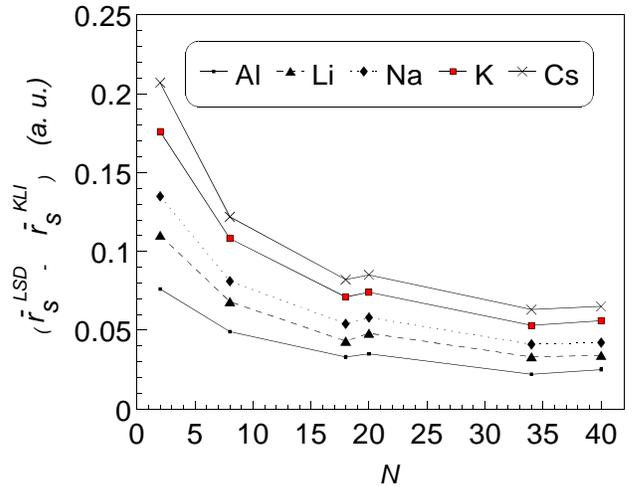}
\caption{\label{fig5} Difference in the equilibrium $r_s$ values
of clusters with different sizes. The difference is larger for
lower-density elements.}
\end{figure}

To illustrate the trend in the $\bar r_s$ values, we plot the
difference $(\bar r_s^{\rm LSD}-\bar r_s^{\rm KLI})$ for all
species in Fig. \ref{fig5}. One notes that for a given element,
the difference is larger for smaller clusters. On the other hand,
the difference for the lower-density element is higher. However,
the difference is about $3\%$ on average. We therefore conclude
that the EEX-SJM-SC predicts smaller bond lengths compared to the
LSDA-SJM-SC. Comparison of the $\bar r_s$ values for the LSDA and
x-LSDA shows that bond lengths in the LSDA is about $1.4\%$ larger
on average. This difference should be attributed to the
correlation effects. On the other hand, the same comparison
between x-LSDA and KLI shows that, except for $N=40$ in Al, $\bar
r_{x-LSDA}>\bar r_{KLI}$ by $1.5\%$ on average. This difference is
due to the self-interaction effects in the Dirac form for the
exchange functional.

Comparison of the equilibrium total energies of the OEP and KLI
shows that OEP energies are on average $0.02\%$ more negative.
This result should be compared to the simple JM
results\cite{Payami05} which is $1.2\%$. On the other hand,
comparison of the exchange energies shows that on the average, the
exchange energies in OEP is $0.34\%$ more negative than those in
the KLI.

In Table \ref{table2}, we have listed the lowest and highest
occupied KS eigenvalues for different schemes. As in the simple JM
\cite{Payami05}, the OEP KS eigenvalue bands are contracted
relative to those of the KLI. That is, for all $N$, the relation
$\Delta^{\rm OEP}<\Delta^{KLI}$ holds. Here,
$\Delta=\varepsilon_H-\varepsilon_L$ is the difference between the
maximum occupied and minimum occupied KS eigenvalues. For the same
external potential, the OEP and KLI results coincide for
two-electron systems and $\Delta=0$. The results in
Table\ref{table2} show that the maximum relative contraction,
$|\Delta^{\rm OEP}-\Delta^{KLI}|/\Delta^{KLI}$, is 2.7$\%$ which
corresponds to Cs$_{18}$.

The same comparisons between OEP and x-LSDA shows that
$E^{OEP}<E^{x-LSDA}$ by $5.2\%$ on average, and
$E_x^{OEP}<E_x^{x-LSDA}$ by $11\%$ on average. The band widths do
not show any regular pattern, however, in the OEP the bands mostly
contract relative to the x-LSDA.

Finally, we compare the results of LSDA and x-LSDA, which will
show the correlation effects. As is seen in Table \ref{table1},
the total energies are close to each other for the high-density
cases. That is, in the high density limit the exchange dominates
the correlation. However, the total energies in the LSDA are more
negative by $10\%$ on average which is due to the correlation
effects. On the other hand, the difference in the exchange
energies is about $0.96\%$ on average which is quite a small
fraction. In the high density limit, the inequality
$E_x^{x-LSD}<E_x^{LSDA}$ holds whereas, in the low density limit
the inequality changes sign.

\begingroup
\squeezetable
\begin{table*}
\caption{\label{table1}Equilibrium sizes, $\bar r_s$, in bohrs,
the absolute values of total and exchange energies in rydbergs are
compared for KLI, OEP,x-LSDA, and LSDA schemes. In the LSDA, the
total energies include the correlation energies as well.}
\begin{ruledtabular}
\begin{tabular}{ccccccccccccccc}
& & &\multicolumn{3}{c}{KLI}&\multicolumn{3}{c}{OEP}&\multicolumn{3}{c}{x-LSDA}&\multicolumn{3}{c}{LSDA}\\
 Atom&$r_s^B$&$N$&$\bar{r}_s$&$-\bar{E}$&$-E_x$&$\bar{r}_s$&$-\bar{E}$&$-E_x$&$\bar{r}_s$&$-\bar{E}$&$-E_x$&
 $\bar{r}_s$&$-\bar{E}$&$-E_x$\\ \hline
Al\footnote{Here, $N$=18 corresponds to Al$_6$ cluster and other
$N$'s do not correspond to a real Al clusters.}
&2.07&2&1.430&1.5700&0.9253&1.430&1.5700&0.9253&1.468&1.4364&0.7574&1.506&1.5585&0.7541 \\
    &&8&1.744&5.8640&3.6018&1.744&5.8647&3.6089&1.775&5.5768&3.2430&1.793&6.1204&3.2361 \\
   &&18&1.876&12.7709&7.9467&1.876&12.7734&7.9760&1.898&12.3315&7.3889&1.909&13.5947&7.3850 \\
   &&20&1.846&14.3309&8.8532&1.847&14.3319&8.8706&1.869&13.8729&8.2870&1.881&15.2718&8.2738 \\
   &&34&1.928&23.9914&14.9857&1.928&23.9968&15.0339&1.944&23.3442&14.1758&1.950&25.7679&14.1829 \\
   &&40&1.901&28.2841&17.5064&1.901&28.2863&17.5348&1.893&27.6468&16.9255&1.926&30.4900&16.7211 \\
&&&&&&&&&&&&&& \\
Li&3.28&2&2.698&1.0076&0.5748&2.698&1.0076&0.5748&2.756&0.9247&0.4710&2.808&1.0264&0.4745 \\
      &&8&2.966&3.9138&2.2501&2.966&3.9144&2.2557&3.013&3.7326&2.0296&3.034&4.1678&2.0363 \\
     &&18&3.086&8.6776&5.0261&3.086&8.6798&5.0506&3.117&8.3934&4.6730&3.129&9.3963&4.6879 \\
     &&20&3.059&9.6670&5.5418&3.059&9.6682&5.5553&3.094&9.3791&5.1972&3.107&10.4905&5.2078 \\
     &&34&3.134&16.3774&9.4868&3.134&16.3823&9.5298&3.157&15.9553&8.9684&3.167&17.8728&8.9866 \\
     &&40&3.111&19.1876&10.9835&3.111&19.1898&11.0052&3.136&18.7942&10.5229&3.145&21.0418&10.5398 \\
&&&&&&&&&&&&&& \\
Na&3.99&2&3.403&0.8409&0.4785&3.403&0.8409&0.4785&3.475&0.7721&0.3918&3.538&0.8646&0.3964 \\
      &&8&3.664&3.2841&1.8579&3.663&3.2846&1.8632&3.719&3.1343&1.6774&3.745&3.5261&1.6856 \\
     &&18&3.784&7.3064&4.1549&3.784&7.3084&4.1772&3.821&7.0700&3.8619&3.838&7.9710&3.8769 \\
     &&20&3.758&8.1240&4.5669&3.758&8.1251&4.5794&3.800&7.8873&4.2867&3.816&8.8856&4.2995 \\
     &&34&3.834&13.7980&7.8340&3.833&13.8028&7.8751&3.862&13.4458&7.4017&3.875&15.1665&7.4223 \\
     &&40&3.813&16.1410&9.0432&3.813&16.1431&9.0632&3.843&15.8198&8.6726&3.855&17.8365&8.6906 \\
&&&&&&&&&&&&&& \\
K&4.96&2&4.354&0.6882&0.3920&4.354&0.6882&0.3920&4.443&0.6321&0.3207&4.530&0.7147&0.3258 \\
     &&8&4.609&2.6951&1.5054&4.609&2.6955&1.5098&4.680&2.5738&1.3597&4.717&2.9204&1.3682 \\
    &&18&4.734&6.0102&3.3659&4.734&6.0121&3.3860&4.781&5.8178&3.1269&4.805&6.6130&3.1427 \\
    &&20&4.710&6.6722&3.6887&4.710&6.6733&3.7002&4.760&6.4820&3.4670&4.784&7.3628&3.4795 \\
    &&34&4.787&11.3534&6.3392&4.787&11.3579&6.3782&4.824&11.0656&5.9836&4.840&12.5823&6.0079 \\
    &&40&4.768&13.2650&7.2975&4.768&13.2671&7.3162&4.808&13.0090&7.0029&4.824&14.7863&7.0226 \\
&&&&&&&&&&&&&& \\
Cs&5.63&2&5.006&0.6123&0.3494&5.006&0.6123&0.3494&5.109&0.5624&0.2856&5.213&0.6395&0.2910 \\
      &&8&5.261&2.3990&1.3322&5.261&2.3994&1.3363&5.342&2.2918&1.2039&5.383&2.6135&1.2133 \\
     &&18&5.390&5.3547&2.9775&5.389&5.3564&2.9963&5.443&5.1842&2.7658&5.472&5.9215&2.7821 \\
     &&20&5.366&5.9403&3.2589&5.366&5.9414&3.2701&5.425&5.7729&3.0640&5.451&6.5894&3.0784 \\
     &&34&5.445&10.1156&5.6044&5.445&10.1200&5.6423&5.488&9.8599&5.2875&5.508&11.2652&5.3122 \\
     &&40&5.428&11.8123&6.4416&5.428&11.8144&6.4598&5.472&11.5881&6.1873&5.493&13.2347&6.2061 \\
\end{tabular}
\end{ruledtabular}
\end{table*}
\endgroup

\begingroup
\squeezetable
\begin{table*}
\caption{\label{table2}The absolute values at equilibrium state of
the highest occupied and lowest occupied Kohn-Sham eigenvalues in
rydbergs are compared for KLI, OEP, x-LSDA, and LSDA schemes. }
\begin{ruledtabular}
\begin{tabular}{cccccccccc}
& & \multicolumn{2}{c}{KLI}&\multicolumn{2}{c}{OEP}&\multicolumn{2}{c}{x-LSDA}&\multicolumn{2}{c}{LSDA}\\
 Atom&$N$&$-\varepsilon_L$&$-\varepsilon_H$&$-\varepsilon_L$&$-\varepsilon_H$&$-\varepsilon_L$
 &$-\varepsilon_H$&$-\varepsilon_L$&$-\varepsilon_H$\\ \hline
Al
 &2&0.8152&0.8152&0.8152&0.8152&0.4367&0.4367&0.5012&0.5012 \\
 &8&1.1142&0.6714&1.1088&0.6713&0.8201&0.3919&0.8821&0.4605 \\
&18&1.1727&0.5507&1.1619&0.5492&0.9497&0.3310&1.0129&0.4009 \\
&20&1.1856&0.5000&1.1804&0.4993&0.9665&0.2964&1.0282&0.3622 \\
&34&1.2055&0.4826&1.1998&0.4789&1.0192&0.2939&1.0853&0.3649 \\
&40&1.2202&0.4490&1.2136&0.4450&1.0541&0.2761&1.0965&0.3401 \\
&&&&&&&&& \\
Li&2&0.4777&0.4777&0.4777&0.4777&0.2445&0.2445&0.2983&0.2983 \\
&8&0.5760&0.4157&0.5735&0.4158&0.3935&0.2383&0.4476&0.2937 \\
&18&0.5935&0.3601&0.5879&0.3591&0.4523&0.2179&0.5062&0.2738 \\
&20&0.5889&0.3221&0.5865&0.3228&0.4522&0.1893&0.5061&0.2423 \\
&34&0.6029&0.3282&0.5991&0.3258&0.4832&0.2055&0.5374&0.2620 \\
&40&0.5979&0.2971&0.5950&0.2960&0.4816&0.1832&0.5355&0.2366 \\
&&&&&&&&& \\
Na&2&0.3883&0.3883&0.3883&0.3883&0.1951&0.1951&0.2437&0.2437 \\
&8&0.4467&0.3406&0.4451&0.3408&0.2963&0.1936&0.3453&0.2434 \\
&18&0.4544&0.2989&0.4502&0.2981&0.3373&0.1805&0.3859&0.2308 \\
&20&0.4485&0.2672&0.4470&0.2682&0.3361&0.1565&0.3851&0.2042 \\
&34&0.4583&0.2750&0.4551&0.2730&0.3588&0.1728&0.4078&0.2236 \\
&40&0.4520&0.2480&0.4502&0.2477&0.3564&0.1534&0.4055&0.2017 \\
&&&&&&&&& \\
K&2&0.3100&0.3100&0.3100&0.3100&0.1526&0.1526&0.1957&0.1957 \\
&8&0.3408&0.2733&0.3396&0.2735&0.2188&0.1536&0.2622&0.1977 \\
&18&0.3416&0.2422&0.3385&0.2415&0.2463&0.1458&0.2894&0.1902 \\
&20&0.3356&0.2169&0.3349&0.2182&0.2450&0.1266&0.2885&0.1687 \\
&34&0.3419&0.2246&0.3393&0.2230&0.2607&0.1413&0.3042&0.1861 \\
&40&0.3355&0.2025&0.3346&0.2027&0.2585&0.1255&0.3023&0.1683 \\
&&&&&&&&& \\
Cs&2&0.2723&0.2723&0.2723&0.2723&0.1324&0.1324&0.1724&0.1724 \\
&8&0.2923&0.2405&0.2913&0.2406&0.1843&0.1343&0.2247&0.1752 \\
&18&0.2907&0.2141&0.2880&0.2134&0.2061&0.1285&0.2461&0.1696 \\
&20&0.2850&0.1921&0.2847&0.1935&0.2048&0.1118&0.2454&0.1509 \\
&34&0.2897&0.1992&0.2873&0.1977&0.2176&0.1252&0.2580&0.1668 \\
&40&0.2835&0.1797&0.2831&0.1801&0.2157&0.1115&0.2564&0.1512 \\
\end{tabular}
\end{ruledtabular}
\end{table*}
\endgroup
\section{Summary and Conclusion}\label{sec4}
In this work, we have considered the exact-exchange stabilized
jellium model with self-compression in which we have used the
exact orbital-dependent exchange functional. This model is applied
for the simple metal clusters of Al, Li, Na, K, and Cs. For the
local exchange potential in the KS equation, we have solved the
OEP integral equation by the iterative method. By finding the
minimum energy of an $N$-atom cluster as a function of $r_s$, we
have obtained the equilibrium sizes and energies of the
closed-shell clusters ($N=2,8,18,20,34,40$) for the four schemes
of LSDA, KLI, OEP, and x-LSDA. The results show that in the
EEX-SJM, the clusters are more contracted relative to the
x-LSDA-SJM, i.e., $1.5\%$ more contraction on average. The KLI and
OEP results show equal values (up to three decimals) for the
equilibrium $r_s$ values. The equiliblium sizes in LSDA and x-LSDA
differ by $1.4\%$ on average. In the LSDA and KLI the difference
in $3\%$ on average. The total energies in the OEP are more
negative than the KLI by $0.02\%$ on the average. It should be
mentioned that in the simple JM the KLI and OEP total energies for
Al were positive (except for $N=2$). On the other hand, the
exchange energies in the OEP is about $0.34\%$ more negative than
that in the KLI. Comparison of the OEP and x-LSDA shows a
duifference of $5.2\%$ in the total energies and $11\%$ in the
exchange. The difference in the exchange energies of LSDA and
x-LSDA is small (about $0.96\%$) whereas the total energy in the
LSDA is about $10\%$ more negative which is due to the correlation
effects. The widths of the occupied bands,
$\varepsilon_H-\varepsilon_L$ in the OEP are contracted relative
to those in the KLI by at most $2.7\%$.

\end{document}